\newcommand\pa{\partial}
\newcommand\beq{\begin{equation}}
\newcommand\eeq{\end{equation}}
\newcommand\beqnl{\begin{eqnarray}}
\newcommand\beqna{\begin{eqnarray*}}
\newcommand\eeqna{\end{eqnarray*}}
\newcommand\eeqnl{\end{eqnarray}}

 \def\NN{\hbox{\sf I\kern-.13em\hbox{N}}}
 \def\HH{\hbox{\sf I\kern-.13em\hbox{H}}}
 \def\DD{\hbox{\sf I\kern-.13em\hbox{D}}}
 \def\RR{\hbox{\sf I\kern-.14em\hbox{R}}}
 \def\CC{\hbox{\sf I\kern-.44em\hbox{C}}}
 \def\ZZ{{\hbox{\sf Z\kern-.43emZ}}}
 \def\QQ{\hbox{\sf C\kern -.48emQ}}
 \def\Cc{\hbox{\sf C\kern -.47em {\raise .48ex \hbox{$\scriptscriptstyle |$}}
   \kern-.5em {\raise .48ex \hbox{$\scriptscriptstyle |$}} }}
 \def\Qq{\hbox{\sf Q\kern -.57em {\raise .48ex \hbox{$\scriptscriptstyle |$}}
   \kern-.55em {\raise .48ex \hbox{$\scriptscriptstyle |$}} }}

\documentclass{ws-ijmpd}
\usepackage{epsfig}
\begin{document}
\vskip1pc
 
\title{Black Holes and the Third Law of Thermodynamics}
\author{F. Belgiorno\footnote{E-mail address: belgiorno@mi.infn.it}}
\address{Dipartimento di Fisica, Universit\`a degli Studi di Milano,\\ 
Via Celoria 16, 20133 Milano, Italy, and\\
I.N.F.N., sezione di Milano, Italy}
\author{M. Martellini\footnote{E-mail address: 
maurizio.martellini@uninsubria.it}}
\address{Dipartimento di Fisica, Universit\`a degli Studi dell'Insubria,\\  
Via Valleggio 11, Como, and\\
I.N.F.N., sezione di Milano, Italy, and\\ 
Landau Network at ``Centro Volta'', Como, Italy}
\maketitle

\begin{abstract}
\vskip -0.3truecm

We discuss in the framework of black hole 
thermodynamics some aspects relative to the 
third law in the case of black holes of the Kerr-Newman family. 
In the light of the standard proof of the equivalence between the 
unattainability of the zero temperature and the entropic version of 
the third law it is remarked that the unattainability has a special 
character in black hole thermodynamics. 
Also the zero temperature limit which obtained in the case of very 
massive black holes is discussed and it is shown that a violation of the 
entropic version in the charged case occurs. 
The violation of the Bekenstein-Hawking law in favour of zero entropy $S_E=0$ 
in the case of extremal black holes is suggested as a natural solution 
for a possible violation of the second law of thermodynamics. 
Thermostatic 
arguments in support of the unattainability are explored, and 
$S_E=0$ for extremal black holes is shown to be again a 
viable solution. 
The third law of black hole dynamics by W.Israel 
is then interpreted as 
a further strong corroboration to the picture of a discontinuity 
between extremal states and non-extremal ones.

\vskip -0.3truecm
\end{abstract}

\vskip0.2cm\noindent
\\PACS:  04.70.Dy, 05.70.-a\\

\vskip2pc

\section{Introduction}
\label{intro}

The third law of thermodynamics is explored in light of 
the correspondences existing between 
standard thermodynamics and black hole thermodynamics 
\cite{BCH,bardeen,davies,israel,wald}. 
We summarize some aspects of the third law in standard 
thermodynamics and also some 
related results in black hole thermodynamics. 
In black hole thermodynamics, the unattainability 
(U) of $T=0$ holds and the entropic version (N) fails; 
it is underlined that (U) has a special status, is clearly 
not equivalent to (U) as it is realized in standard thermodynamics. 
Moreover, the limit 
$T\to 0^+$ obtained for black hole mass $M\to \infty$ is  
explored. (N) is shown to hold in the case of uncharged black 
holes; at the same time, (U) holds also for the charged case. 
Then we focus our attention on extremal states. 
An analysis in a thermostatic framework suggests that it is appealing  
to abandon the Bekenstein-Hawking law in the case of the 
extremal black holes and to assume that extremal black hole 
have zero entropy $S_E=0$.    
Thermostatic reasons for the unattainability (U) of extremal 
states are then analyzed; in particular, 
if $S_E=0$, then it is possible to 
corroborate the unattainability also in a thermostatic 
framework. Unattainability would mean simply the impossibility 
of a process violating the second law of thermodynamics. 
The analysis of the implications of Israel's proof \cite{israel}, 
to be interpreted in a irreversible thermodynamic framework, 
corroborates the statement that 
extremal states have to be discontinuous with respect to  
the equilibrium thermodynamic space of non--extremal ones.\\ 
\\
\noindent The plan of the paper is the following.  
In sect. \ref{law} a review of discussions and results about third law of 
thermodynamics in black hole physics 
\cite{BCH,bardeen,davies,israel,wald,racz} 
and of results about extremal black hole  entropy 
\cite{hawross,hawsurf,vanzo,kiefer}; 
a short review of the status of the third law in standard thermodynamics 
\cite{simon,zemansky,landsberg,landrmp,landstat} and Nernst's  
theorem is made in sect. \ref{nernstheat}. 
In sect. \ref{bhbranches} an analysis of the 
limit $T\to 0^+$ in black hole thermodynamics follows. A first analysis 
concerns extremal states and then the limit $T\to 0^+$ associated 
with infinitely massive black holes is explored.\\
In sect. \ref{ebh} the thermodynamic properties of 
extremal black holes are enhanced.  
In sect. \ref{foliation} Carath\'eodory's approach to black hole 
thermodynamics is applied in order to study the properties 
of the extremal surface $T=0$ and it is shown that, by including it 
in the thermodynamic manifold, one cannot obtain a well-behaved 
thermodynamic foliation. $S_E=0$ is proposed 
as a solution to this problem and the introduction of this discontinuity 
between non-extremal states and extremal ones is shown to be viable.  
In sect. \ref{thermouna} unattainability is discussed in a 
thermostatic framework. 
In sect. \ref{irre} a translation of Israel's result \cite{israel} 
in a irreversible thermodynamic frame is made. 
Appendices A,B,C and D concern further aspects of the 
physics involved in our paper.

\section{The third law}
\label{law}

In standard thermodynamics there are two formulations 
of the third law. 
The entropic version of Nernst's theorem (N) 
states that, for every system, if one considers the 
entropy as a function of the temperature $T$ and of other 
macroscopic parameters $x^1,\ldots,x^n$, the entropy 
difference 
$\Delta_T S\equiv S(T,x^1,\ldots,x^n)-S(T,\bar{x}^1,\ldots,\bar{x}^n)$ 
goes to zero as $T\to 0^+$
\beq
\lim_{T\to 0^{+}} \Delta_T S =0
\label{nland}
\eeq
for any choice of $(x^1,\ldots,x^n)$ and of  $(\bar{x}^1,\ldots,\bar{x}^n)$.  
Thus (N) requires that 
the limit $\lim_{T\to 0^{+}} S(T,x^1,\ldots,x^n)$ is a constant 
$S_0$ which does not depend on the macroscopic parameters $x^1,\ldots,x^n$. 
Planck's restatement of Nernst's 
postulate fixes the entropy constant $S_0$ at $T=0$ to be zero
\footnote{This is mandatory in the case of homogeneous thermodynamics, 
as is shown in \cite{belg30} and it is trivial to prove.}.  
Sometimes, the (N) version is expressed by saying that the zero 
temperature states of a system are isentropic. The latter 
statement is at least ambiguous, if the entropy is allowed 
to be discontinuous; the statement involving the limit of $S$ 
as $T\to 0^+$ is in any case to be preferred.      
The unattainability version (U) can be expressed as the impossibility 
to reach the absolute zero of the temperature by means of a 
finite number of thermodynamic processes. Both the above formulations 
are due to Nernst. It is 
generally assumed that the two formulations are equivalent. 
Actually, this equivalence is not automatic, as it results 
from a  discussion in   \cite{landsberg,landrmp,landstat}. We  
mean to come back to this topic in sect. \ref{ebh}, 
where its relevance in black 
hole thermodynamics is enhanced.\\

The third law of thermodynamics in black hole physics has been 
discussed since the formulation of the laws of black hole 
mechanics \cite{BCH}. In fact, in   \cite{BCH} the analogy between 
the standard third law, in the form of unattainability (U) of the 
absolute zero temperature, and the unattainability of the 
extremal states by means of a finite number of physical processes 
is remarked \cite{bardeen}. A more recent result about the 
unattainability is found in   \cite{israel}, where the unattainability 
is rigorously obtained under suitable hypotheses. We discuss 
in the following this result further on. (N) is explored in the framework 
of black hole thermodynamics e.g. in   \cite{davies}. 
Therein it is stressed the 
failure of the entropic side of the third law in black hole 
thermodynamics. 
On the side of (N), we recall also some results obtained in the 
framework of gravitational partition function calculations. 
In   \cite{hawross,hawsurf} the entropy of an extremal  
Reissner--Nordstr\"{o}m black hole is predicted to be zero 
and this result is related with the boundary structure of the 
spacetime. Analogous statements are found in   \cite{vanzo} and 
a further corroboration of $S=0$ for extremal 
Reissner--Nordstr\"{o}m black holes  
appears in   \cite{kiefer}, where a semiclassical calculation 
of the entropy in canonical quantum gravity is made. The latter approach 
leads to a result that introduces a violation of the 
Bekenstein--Hawking law and is in agreement with the requirement 
of isentropic zero temperature states. This 
isentropy is not equivalent to (N), even if it seems to match 
Planck's requirement of zero entropy for any system at the 
absolute zero of the temperature. A deeper discussion is  
found in the following sections.  
On the other hand, superstring theory and supergravity allow 
again the opposite result in which $S=A/4$ for extremal black 
holes. We don't discuss herein the latter 
approach.\\ 

Doubts about the validity of thermodynamics 
for values of $T$ very near the absolute zero have been raised, 
when finite-size systems are taken into account. 
A thermodynamic description of a ``standard" system below 
a given temperature is impossible according to Planck,  
because of a reduction of the effective degrees of freedom 
making impossible even to define an entropy. 
Only statistical mechanics is then viable. In 
  \cite{munster} this breakdown of thermodynamics 
near the absolute zero is shown to occur because of 
finite size effects, which make impossible to neglect 
statistical fluctuations 
in the calculation of thermodynamic quantities like e.g. $T,S$. 
On the black hole side of this topics, arguments that are in some 
sense of the same nature as Planck's ones are found in 
  \cite{preskill}. In fact, therein hints  
against the possibility of a thermal description of near extremal states, 
because of the occurrence of uncontrollable thermodynamic fluctuations, 
are given, and are related to the finite size nature of black holes 
(note also that a notion of thermodynamic limit is missing in the black 
hole case).

We also limit ourselves to point out that 
an unconventional discussion of black hole thermodynamics 
in a non-equilibrium framework and a unconventional 
discussion also of the third law are found 
in \cite{nieu}.

Herein, we discuss the extremal limit of black hole 
thermodynamics, and also the third principle in black hole 
thermodynamics with respect 
to the limit $T\to 0^+$ obtained for infinite black hole mass 
$M\to \infty$. 

\section{The equivalence (U)$\Leftrightarrow$(N) revisited. Landsberg's 
analysis}
\label{nernstheat}

In the following, 
we first focus our attention on the relation 
between (N) and (U) and on the possibility to de--link 
the unattainability from the entropic version of Nernst's theorem. 
The double implication (U)$\Leftrightarrow$(N), 
according to the analysis developed by Landsberg in   
\cite{landsberg,landstat}, 
relies on some hypotheses that it is interesting to recall. 
In the following sections we discuss some properties of 
black hole thermodynamics near the absolute zero in the light 
of this analysis.  
For a detailed discussion about the third law in standard thermodynamics 
see also \cite{belg30,belg31}. 

\subsection{(U)$\Rightarrow$(N)}

\begin{figure}[h]
\setlength{\unitlength}{1.0mm}
\centerline{\epsfig{figure=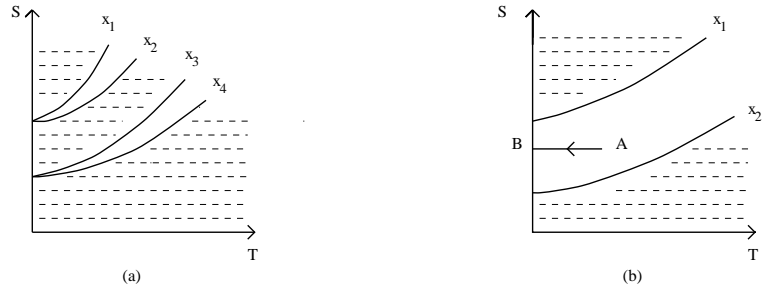,height=10cm,angle=-90}}
\vspace{0.2cm}
\caption{ (a) Multi--branches structure of the thermodynamic space. 
According to Landsberg, 
it implies the validity of (U) and the 
violation of (N). (b) Violation of (N) that implies a violation 
of (U), due to the presence of the isentropic AB. Landsberg conjectures 
that (U)  
holds if a discontinuity near $T=0$ occurs. See also the text.  
In (a) and (b) the dashed regions are forbidden. }
\label{fig1}
\end{figure}

A detailed analysis shows that in standard thermodynamics 
unattainability (U) implies (N) if the following conditions 
a), b), c) are 
satisfied \cite{landsberg,landstat}:\\ 
\\
a) The stability 
condition  
$(\partial S/\partial T)_{x}>0$ is satisfied for any 
transformation such that the external parameters 
(or deformation coordinates; in our discussion we include 
constitutive coordinates in the set of deformation coordinates), 
collectively indicated by means of $x$, are kept fixed; 
these transformations are called isometric 
transformations \cite{buchdahl}. As a consequence, the 
heat capacity $C_{x}$ at constant deformation parameters $x$ has to be 
positive $\forall\; T>0$.\\
\\ 
This hypothesis is in general ensured by the 
convexity/concavity properties of the thermodynamic potentials; 
as a consequence, in Landsberg's works 
\cite{landsberg,landrmp,landstat} 
this hypothesis is actually assumed to be always satisfied, 
so it is not considered as a possible cause of failure of the 
double implication (U)$\Leftrightarrow$(N). Given also 
the peculiar thermodynamic properties of black holes, 
we must choose a) as a further hypothesis to be discussed.\\ 
\\
b) There are no multiple branches in thermodynamic configuration 
space.\\
\\
This condition is introduced in order to avoid some 
pathological situations discussed in   \cite{landsberg,landrmp} 
(no physical behavior corresponds to them; see fig. \ref{fig1}).\\
\\
c) There is no discontinuity in thermodynamic properties of the 
system near the absolute zero.\\
\\
In   \cite{landsberg} a careful discussion of the conditions to 
be satisfied in order to ensure (U) is contained. In particular, 
by following   \cite{landsberg}, if a),b),c) hold and moreover 
(N) fails, then $T=0$ is attainable. 
If a),b) and c) hold, then (U) implies 
(N). If a),b) hold and (N) fails, then (U) implies that a discontinuity 
near the absolute zero has to occur, and such a discontinuity has 
to prevent the attainability of $T=0$ (violation of c)) \cite{landsberg}. 
Anyway, in standard thermodynamics 
a violation of c) is ruled out \cite{landsberg}, and 
(U) is associated with the impossibility to 
get states at $T>0$ isentropic to states at $T=0$.\\

A further condition ``entropies don't diverge as $T\to 0^+$" 
is also introduced in   \cite{landsberg,landstat} in order to take 
into account the actual 
behavior near zero temperature of the standard thermodynamic 
systems. In fact, a priori, (N) could hold also if one should find a 
divergence in the entropy as $T\to 0^+$ not depending on $x$ 
\cite{landsberg}. However, we don't introduce this further 
hypothesis in view of our analysis in sect. \ref{infinitm}. 
 
About possible failures of the implication 
(U)$\Rightarrow$(N) see also  \cite{haase}.
 
As far as black hole thermodynamics is concerned, we note that the 
violation of (N) is such that (U) cannot be interpreted 
as absence of isentropic transformations which allow to reach $T=0$.  
See sect. \ref{foliation}

\subsection{(N)$\Rightarrow$(U)}
\label{uenne}

A full implication (N)$\Rightarrow$(U) is possible  
in the case of thermodynamic processes which consist of an 
alternate sequence of quasi-static adiabatic transformations 
and quasi-static isothermal transformations (class P(x) according to 
  \cite{landsberg,landrmp}). 
Actually, a more general notion of unattainability can be 
assumed, that is, ``zero temperature states don't 
occur in the specification of attainable states of 
systems". This is almost literally the (U4) principle introduced  
by Landsberg  \cite{landsberg,landrmp}. (U4)-unattainability 
states that no process allows 
to reach states at $T=0$, even 
as transient non-equilibrium states. 
Then (N) can fail and (U) can still be valid. In general, the 
latter hypothesis allows a de--linking of (U) and (N) and implies that (N)
$\not \Rightarrow$(U) and (U)$\not \Rightarrow$(N) \cite{landsberg,landrmp}. 
But such a de--linking occurs under particular conditions. The 
failure of the implication (U)$\Rightarrow$(N) requires 
again a rejection of one of the hypotheses b),c) 
above, whereas 
(N)$\Rightarrow$(U) fails if processes not belonging 
to the aforementioned class P(x) allow 
to reach $T=0$ \cite{landsberg,landrmp}. 

We recall that 
the standard approach to Nernst's theorem  
involves heat capacities and runs e.g. as in 
\cite{munster2,guggenheim}. We further limit ourselves to 
recall that black hole 
thermodynamics has some very peculiar features that make it 
special with respect to ``standard" thermodynamics ( 
see e.g. \cite{landblack,martinez}).

\section{Black holes branches as $T\to 0^+$}
\label{bhbranches}

In this section we discuss black hole thermodynamic branches 
near $T=0$. We first recall that for the entropy, as a function of 
$T,M,Q$, there is a branching into two different functions.  
In fact, although the entropy is a continuous function 
of $M,J,Q$, there are points such that the state equation  
$T(M,J,Q)$ cannot be inverted in order to get $M(T,J,Q)$.  
It happens that $\pa T/\pa M =0$ can be satisfied on 
suitable submanifolds, where standard conditions for 
the implicit function theorem fail. As a consequence, 
one can invert $T$ and obtain $M(T,J,Q)$, 
to be substituted into $S(M,J,Q)$ only 
away from these submanifolds, on the two branches 
$\pa T/\pa M >0$ and $\pa T/\pa M <0$,  
which are associated with the zero temperature limits 
obtained as finite mass extremal limit 
and as infinite mass limit respectively. In particular, 
for the same value of the variables $T,J,Q$ it is possible to 
get two different values of $S$. 
This is  a sufficient reason for a multi-branching 
in the $S-T$ plane. These branches describe 
different systems.\footnote{Points where $\pa T/\pa M =0$ correspond 
to critical submanifold 
points where $C_{QJ}$ diverges and changes sign. 
It has been proposed that a second order phase transition 
takes place there \cite{davies}.}  
We first consider the finite mass extremal limit and then 
the infinite mass limit. 

\subsection{Black hole extremal limit $M<+\infty$}
\label{extrbranch}

\begin{figure}[h]
\setlength{\unitlength}{1.0mm}
\centerline{\psfig{figure=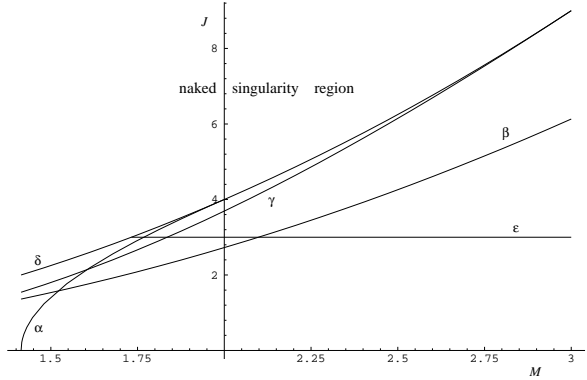,height=5cm,angle=0}}
\vspace{0.2cm}
\caption{In the Kerr case, the isentrope $\alpha:\; J=\sqrt{8\; M^2 - 16}$,  
the line $\epsilon:\; J=$ const.=3, and the line $\gamma:\; 
J=(6\; M^3)/(M^2+9)  $, 
which corresponds to $\Omega=$ const. =1/6, are shown. Extremal states 
lie on the line $\delta:\; J=M^2$, above which naked solutions are 
found. The isentrope $\alpha $ ends at $M=2$, where it is tangent to 
the extremal manifold (see also Appendix A).   
Above the line $\beta:\; J=\sqrt{2\sqrt{3}-3}\; M^2  $ the 
heat capacity $C_J$ is positive. The line $ J=(6\; M^3)/(M^2+9)$ intersects 
the extremal manifold for $M=3$ (tangent to it).}
\label{kekk1}
\end{figure}

The violation of the third law in black hole thermodynamics 
as the extremal boundary is approached is well-known \cite{davies}. 
For the sake of completeness, we show that 
(\ref{nland}) fails in the general case of a Kerr--Newman black hole.  
Let us define 
\beqna
M^2_E&\equiv&\frac{1}{2} (Q^2+\sqrt{Q^4+4J^2})\\
M^2_N&\equiv&\frac{1}{2} (Q^2-\sqrt{Q^4+4J^2})<0;
\eeqna
$M_E^2,M_N^2$ are the roots of the equation $(M^2)^2-Q^2 M^2-J^2=0$ and 
$M_E^2$ corresponds to the squared mass of the extremal 
Kerr--Newman solution having charge $Q$ and angular momentum $J$. 
It is useful to explicit the following relations between the 
above roots and the charge $Q$ and the angular momentum $J$ of the 
black hole: $Q^2=M^2_E+M^2_N$, $J^2=- M^2_E  M^2_N$. Moreover, 
the difference $(M^2_E-M^2_N)$ is related to the area of the extremal 
solution for given values of $Q,J$ by $A_E=4 \pi (M^2_E-M^2_N)$. 
We can rewrite 
\beqna
T&=&\frac{M}{2 \pi} 
\frac{\left( (M^2-M^2_E)(M^2-M^2_N) \right)^{1/2}}
{\left( M^2+\left((M^2-M^2_E)(M^2-M^2_N)\right)^{1/2}\right)^2
-M^2_E M^2_N}\cr
S&=&\frac{\pi}{M^2} 
\left( \left( M^2+\left((M^2-M^2_E)(M^2-M^2_N)\right)^{1/2}\right)^2
-M^2_E M^2_N \right).
\eeqna
It is easy to show that, 
near the extremal states $M^2\sim M_E^2$, one has $
M^2=M_E^2+4 \pi^2 M_E^2 (M^2_E-M^2_N) T^2+\cdots$. 
Then, for $T\to 0^+$, one finds $S(T,Q,J)\sim \pi  (M_E^2-M_N^2)+ 4 
\pi^2  M_E (M_E^2-M_N^2) \;T+\cdots$, thus 
\beqnl 
\lim_{T\to 0^+} \left( S(T,Q_1,J_1)-S(T,Q_2,J_2)\right) &=&
 \frac{1}{4} (A_{1E}-A_{2E})\cr
&=&
\pi ( \sqrt{{Q_1}^4+4 {J_1}^2}- \sqrt{{Q_2}^4+4 {J_2}^2} ).
\label{limext}
\eeqnl
As expected, the difference in entropies is proportional to the 
difference of the areas of the corresponding extremal solutions, which 
depend on the macroscopic parameters. 
The limit (\ref{limext}) is to be intended as a right limit 
as $T\to 0^{+}$. 
The discussion which is 
developed in sect. \ref{foliation} shows that 
it could be physically improper to assign by continuity a value 
to the entropy on the boundary $T=0$ of the thermodynamic manifold,  
i.e., it could be improper to assign the value $S_E=A_E/4$ to 
extremal states.\\
The failure of (N) implies that (U) cannot be intended as 
absence of adiabatic transformations reaching $T=0$. It is interesting to 
notice that concavity (hypothesis a)) fails, as it is well-known, in 
black hole thermodynamics \cite{tranah}.  In the black hole case, 
there exist 
curves approaching $T=0$ such that $C_x>0$ and other such that 
$C_y<0$. The existence of 
paths with $C_x<0$ allowing to approach $T=0$ is evident in the 
Kerr case \cite{davies}, where $C_{\Omega}<0$  and $C_{J}>0$ 
near the extremal limit. Cf. fig. \ref{kekk1}, where the corresponding 
curves in the thermodynamic domain are shown. 

In the general Kerr--Newman case,
near the extremal state the heat capacity 
$C_{J,Q}$ is positive and goes to zero as $T$ 
at the extremal limit\cite{davies}. 
Other heat capacities at constant deformation parameters can 
be taken into account \cite{tranah}, 
(e.g. 
$C_{\Phi,Q}=C_{J,\Omega}; C_{\Omega,Q}, 
C_{J,\Phi}, C_{\Omega,\Phi}$). Some of them can be negative near $T=0$. 
Note that, however, the presence of paths (in the thermodynamic space) with 
negative heat capacity is in 
general not sufficient to ensure the violation of (N). It only allows 
to de-link the violation of (N) from the validity of (U), in fact 
the non-uniformity of the sign of heat capacities near $T=0$ can 
allow adiabatic paths reaching $T=0$. See also 
Appendix C and \cite{belg31}.

\subsection{The branch $T\to 0^+$ for $M\to \infty$}
\label{infinitm}

\begin{figure}[h]
\setlength{\unitlength}{1.0mm}
\centerline{\psfig{figure=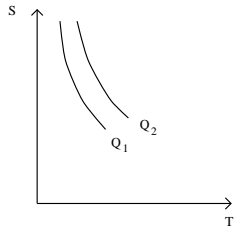,height=3cm,angle=-90}}
\vspace{0.2cm}
\caption{Violation of (N) in the large--$M$ limit.  The qualitative 
behavior of the entropy is displayed for two different value of the black hole 
charge. Two isentropic lines are qualitatively displayed.}
\label{fig2}
\end{figure}

In black hole thermodynamics another limit 
of zero temperature is sometimes considered \cite{davies,lousto}.  
It is the limit as 
$M\to \infty$ e.g. in the Schwarzschild case. In fact, $T\sim 0$ only near the 
extremal 
states or for very large masses. 
But the latter limit cannot 
be considered on the same footing as the limit where  
extremal states are approached, 
indeed it is physically related to an unattainability 
principle in a straightforward way.  
No infinite mass can be allowed on physical grounds, whereas no 
hindrance to consider e.g. $Q^2=M^2$ in the Reissner--Nordstr\"{o}m 
case is a priori given. However, an astrophysical black hole represents 
from a thermodynamic point of view a system reaching temperatures even 
much lower than the ones involved in experiments of low temperature 
physics and in actual experimental validation of (N) and, moreover, 
it is interesting to stress that black hole 
thermodynamics allows to get systems having a very low temperature 
and a huge entropy in contrast with the low temperature behavior of 
standard systems. In particular, Planck's postulate $S\to 0$ for 
$T\to 0^+$ is to some extent maximally violated\footnote{We remark that 
the violation of (N) in presence of diverging deformation parameters 
can occur in 
the case of systems for which (N) holds at finite deformation parameters. 
See also \cite{belg31}.}.  
It is also 
interesting to show that (N) is violated in the sense that 
(\ref{nland}) fails in the case of a charged black hole. 
Of course, such a violation is impossible in the 
case of a Schwarzschild black hole, due to its too constrained thermodynamic 
phase space. Instead, let us first consider the 
Reissner--Nordstr\"{o}m case; in the 
limit $M\to +\infty$ one can invert explicitly the relation 
between $M$ and $T$
$$
M^2\sim \frac{1}{64 \pi^2 T^2}-8 \pi^2  Q^4  T^2;
$$
then 
\beq 
S(T,Q)\sim \frac{1}{16 \pi T^2}-2 \pi  Q^2  + O(T^2) 
\label{cha1}
\eeq
and
\beq
S(T,Q_{1})-S(T,Q_{2})\to 2 \pi \left(Q^2_{2}-Q^2_{1} \right).
\label{cha2}
\eeq
See fig. \ref{fig2}. 
Obviously $S\to \infty$. If one considers the heat capacity $C_Q$ 
corresponding to the process under consideration, one finds that 
$C_Q<0$ so that the process involves the thermally unstable branch of 
black hole thermodynamics. 
We have just shown that (N) fails but for a self--evident 
reason a sort of (U) is automatically ensured. We stress again that 
(U) is not to be intended as the impossibility to reach $T=0$ in a 
finite number of processes in this case, but simply as the 
impossibility to get an infinitely massive black hole.  
(U) amounts to (U4), the most general notion of 
unattainability of the absolute zero. 

The general Kerr--Newman case can be treated analogously. The starting point 
is a large mass expansion of the equation of state for $T$
$$
T=\frac{1}{8\pi} \frac{1}{M}-\frac{1}{128\pi} (4 J^2+Q^4) \frac{1}{M^5}-
\frac{1}{128\pi} Q^2 (4 J^2+Q^4) \frac{1}{M^7}+O\left(\frac{1}{M^9}
\right);
$$
by inverting one finds
$$
\frac{1}{M}=8 \pi T\;+2048 \pi^5 (4 J^2+Q^4) T^5+ 131072 \pi^7 
Q^2 (4 J^2+Q^4) T^7+O\left(T^9\right)
$$
and
\beqnl
S(T,Q_1,J_1)-S(T,Q_2,J_2)&=& 2 \pi \left(Q^2_{2}-Q^2_{1}\right)\cr
&+&48 \pi^3 \left(4 \left(J_2^2-J_1^2\right)+\left(Q^4_2-Q^4_1\right)
\right) T^2+O\left(T^4\right).
\eeqnl
We can deduce that, if the black hole is uncharged but rotating, 
(N) is satisfied because $S(T,J_1)-S(T,J_2)\sim 192 \pi^3  
(J_2^2-J_1^2) T^2 \to 0$ for $T\to 0^+$. So there is evidence in favor of 
the validity of (N) in the case of uncharged rotating black holes of the 
Kerr family on the thermally unstable large mass branch of black hole 
thermodynamics 
(again, $C_J<0$ for $M\to \infty$). Only the presence of the 
charge is actually associated with the failure of 
(N) on this branch. Thermal instability is verified also 
in the general Kerr--Newman case
$$
C_{J,Q}\sim -\frac{1}{8 \pi T^2}-96 \pi^3 (4 J^2+Q^4) T^2+\cdots <0. 
$$
We find a behavior that is 
remarkable also from a thermodynamic point of view. In fact, the validity of 
(U4) can give rise both to (N) and to the failure of (N) \cite{landrmp}.  
States at $T=0$ on this branch of black hole thermodynamics 
are evidently unphysical and disconnected from the 
finite mass states, we have a discontinuity which 
is directly related to the unavailability of an infinite energy which would  
be necessary in this case in order to obtain a zero temperature state (and 
an infinite entropy state). 
\footnote{
It is also remarkable that, if in general 
the entropy diverges for $T\to 0^+$, 
the axis $T=0$ acts as a vertical asymptote for the graph of $S$ in 
the $S-T$ plane and the possibility to find an adiabat reaching $T=0$ 
evidently is missing. The divergence of the entropy as 
$T\to 0^{+}$ suggests that (U) holds (an isentropic 
at $S=+\infty$ appears to be unphysical). 
$T=0$ has to be excluded 
from the physical domain of $S$.  
Moreover, if this infinite entropy in the zero temperature 
limit can be obtained 
at the cost of an infinite energy, as in the black hole case, 
then there is also a (somehow trivial) physical reason for (U).} 

It is also possible to find thermodynamic transformations joining together 
the extremal limit and the large mass limit to the zero temperature. One can 
choose e.g. 
$$
J^2=M^2 (M^2-Q_0^2) \tanh^2 (\frac{M}{M_0})                               
$$
where $M_0, Q_0$ are constants. It is evident that the extremal limit 
is implemented as $M\to \infty$. In general, the extremal limit can be 
approached only asymptotically along these transformations 
and (U) is preserved as above.\\ 
 
Note that, even in the charged case, one finds
$$
\lim_{T\to 0^+} \frac{S(T,Q_1)}{S(T,Q_2)}=1.
$$
(N) in some sense holds at the leading order 
but fails in the charged case when the difference is taken because 
of sub-leading terms depending on the charge $Q$, as it can be 
inferred from (\ref{cha1}),(\ref{cha2}).

\section{Extremal black holes}
\label{ebh}

We shortly re-analyze the status of extremal black holes in the 
framework of black hole thermodynamics. The following considerations 
don't depend on taking a limit as $T\to 0^+$ of the Smarr formula \cite{smarr}, 
even if they are consistent with such a limit. Our point is that, 
in the case a discontinuity between non-extremal and extremal states 
occurs, such a limit makes no sense. Then, can be useful to 
re-analyze the extremal black holes from this point of view.  
Extremal black holes belonging to the Kerr-Newman family are 
characterized by the extremality constraint 
\beq
M^2=Q^2+\frac{J^2}{M^2}.
\label{conextr}
\eeq
In the case of extremal black holes, the constraint equation 
(\ref{conextr}) is equivalent to the (known) fundamental relation
specifying the black hole state:
\beq
M(Q,J)=\frac{1}{\sqrt{2}} \sqrt{Q^2+\sqrt{Q^4+4 J^2}}=M_E.
\label{fundex}
\eeq
$M(Q,J)$ is a quasi-homogeneous function of 
degree $\frac{1}{2}$ and weights $(1/2,1)$ (see \cite{belcar,qotd} for a 
definition of quasi-homogeneous function):
$$
\frac{1}{2} \frac{\partial M}{\partial Q} Q + 
\frac{\partial M}{\partial J} J=\frac{1}{2} M=
\frac{1}{2} \Phi^{\mathrm extr}_{\mathrm bh} Q+
\Omega^{\mathrm extr}_{\mathrm bh} J, 
$$
where $\Omega^{\mathrm extr}_{\mathrm bh},\Phi^{\mathrm extr}_{\mathrm bh}$ 
are the extremal black hole angular 
velocity and electric potential respectively. This quasi-homogeneity 
property is shared with Smarr formula $M(S,Q,J)$ for non-extremal 
black holes [in the latter case, one has a quasi-homogeneous function of 
degree 1/2 and weights $(1,1/2,1)$]. 
By differentiating (\ref{fundex}) it is easy to show that 
along extremal states 
$$
dM=\Omega^{\mathrm extr}_{\mathrm bh} dJ+\Phi^{\mathrm extr}_{\mathrm bh} dQ
$$
This means that the extremal submanifold is an integral manifold for the 
Pfaffian form $\delta Q_{rev}\equiv 
dM - \Omega_{\mathrm bh}\ dJ-\Phi_{\mathrm bh}\ dQ$, i.e., it is 
an adiabatic submanifold. This rephrasing 
is important for the discussion which follows in sect. \ref{foliation}.\\ 

It is also interesting to note that 
the area for an extremal black 
hole can be expressed as 
\beq
A_{\mathrm extr}=4 \pi (r^2_{+}+\frac{J^2}{M^2})_{\mathrm extr}=
4\pi \sqrt{Q^4+4 J^2}.
\label{arextr}
\eeq
From (\ref{arextr}) one gets 
\beq
(dA)_{\mathrm extr} =\frac{8 \pi}{\sqrt{Q^4+ 4 J^2}}(Q^3\ dQ+2J\ 
dJ)=\frac{32 \pi^2}{A} (Q^3\ dQ+2J\ dJ).
\label{daxtr}
\eeq
The extremality constraint does not implies that $(dA)_{\mathrm extr}$ 
vanishes, and along extremal states in the case of 
Reissner--Nordstr\"{o}m and Kerr black holes 
it is impossible to get $dA_{\mathrm extr}=0$ (i.e. the equation 
$A_{\mathrm extr}=$ const. is satisfied only for  
$M^2=Q^2=$ const. and $M^4=J^2=$ const. respectively), 
whereas for non--extremal black holes it is possible 
to get solutions for $dA=0$. Solutions of $dA_{\mathrm extr}=0$ are 
instead allowed in the Kerr-Newman case. 
See also Appendix A herein.

\section{Carath\'eodory's approach and the surface $T=0$.} 
\label{foliation}

The approach of Carath\'eodory to 
thermodynamics identifies the 
the infinitesimal heat exchanged reversibly $\delta Q_{rev}$ 
with an integrable Pfaffian form.  
See also  
\cite{belhom}, where an original approach  based on homogeneity symmetry
is developed for standard thermodynamics \cite{belhom}, and 
\cite{belcar,qotd}, where quasi-homogeneity symmetry replaces homogeneity 
in the case of black holes and of some self-gravitating matter systems. 
The Pfaffian form for black holes of the Kerr-Newman family is 
$\delta Q_{rev}\equiv dM-\Omega dJ-\Phi dQ$ and it is 
a non-singular integrable Pfaffian form defining a foliation 
of the thermodynamic manifold by means of the solutions of 
the Pfaffian equation $\delta Q_{rev}=0$. 
This one-form is smooth on the non-extremal submanifold, and  
it is continuous everywhere. Its integrability means that, in the 
inner part of the thermodynamic manifold (non-extremal states)  
$\delta Q_{rev}\wedge d(\delta Q_{rev})=0$ is verified \cite{belcar}. 
Carath\'eodory's approach to thermodynamics 
allows also to understand better 
the status of the surface $T=0$ both in standard thermodynamics 
\cite{belg30} and in black hole thermodynamics. 
We limit ourselves to discuss the latter aspect herein. 
It is known that, according to Frobenius theorem, 
with a suitably regular integrable Pfaffian form is associated a foliation 
of the manifold into disconnected codimension one integral 
submanifolds. If one excludes the extremal boundary from the thermodynamic 
manifold, one finds that the leaves associated with $\delta Q_{rev}$ 
are the manifolds $A=$ const., i.e., $S=$ const. On this restricted manifold, 
the adiabatic inaccessibility property holds, which means that in the 
neighborhood of any point P there exists an infinite number of points 
which cannot be reached from P along solution curves of $\delta Q_{rev}=0$ 
[this property, introduced by Carath\'eodory (see \cite{landstat}), 
is equivalent to the integrability property]. Then, 
for $T>0$ the integral manifolds of the Pfaffian form $\delta Q_{rev}$ 
are the surfaces $S=$ const. Given any non-extremal state, 
any path solving the equation $\delta Q_{rev}=0$ 
in the thermodynamic manifold has to lie on a isentropic 
surface. \\ 
The  extremal submanifold is  very 
peculiar. In fact, the surface $T=0$, which 
corresponds to the extremal submanifold, is an integral manifold 
of the Pfaffian form $\delta Q_{rev}$, in the sense that 
it solves the equation $\delta Q_{rev}=0$, as we have seen in sect. 
\ref{ebh}. It could be considered naively as a leaf, 
but the lack of some regularity 
properties of $\delta Q_{rev}$ on the extremal submanifold 
has important consequences. Let us consider the Reissner-Nordstr\"{o}m 
case. By posing $M^2=x; Q^2=y$ one finds 
\beq
\delta Q_{rev}=\frac{1}{2\sqrt{x}} dx-
\frac{1}{2(\sqrt{x}+\sqrt{x-y})} dy,
\eeq
where $y\leq x$. Given a black hole state $(x_0,y_0)$, 
the states which are adiabatically reachable from it lie on the 
curves that are solutions of the following Cauchy problem
\beqnl
&&\frac{dy}{dx}=1+\sqrt{1-\frac{y}{x}}\\ 
&&y(x_0)=y_0.
\eeqnl
The solution of this problem exists and it is unique 
for any initial non-extremal state; moreover, it corresponds to the 
standard isoareal\footnote{The adjective ``isoareal'' is coined in 
agreement with the standard coining of adjectives in thermodynamics 
(see e.g. isothermal,isochoric,...). An isoareal transformation is 
a transformation in which each state has the same area: 
``isoareal = with the same area''.} solution. 
If, instead, one considers 
an extremal state as initial point, the Cauchy problem 
\beqnl
&&\frac{dy}{dx}=1+\sqrt{1-\frac{y}{x}}\\ 
&&y(x_0)=x_0
\label{extcauchy}
\eeqnl
allows {\sl two solutions}: 
\beq
y(x)=x,
\label{extone}
\eeq
which means that the extremal states are adiabatically 
connected each other, and the solution 
\beq
y(x)=2 \sqrt{x_0} \sqrt{x}-x_0
\label{nonextone}
\eeq
which holds for $x\in (x_0/4,x_0]$ and means that 
extremal states are also adiabatically connected to 
non-extremal ones. The key-point is that on the extremal 
manifold, the right member of the differential equation 
(\ref{extcauchy}) is no more smooth (actually, it is not 
$C^1$ and even the weaker Lipschitz condition is not satisfied). 
\footnote{Without  solutions like 
(\ref{nonextone}), one could conclude that the extremal 
manifold is a leaf of the thermodynamic manifold. 
One is instead forced to introduce a discontinuity in order to 
obtain a well-behaved foliation of the whole thermodynamic manifold.}   
This is a serious problem from a thermodynamic point of view, 
because the adiabatic inaccessibility is jeopardized  
by the $T=0$ manifold. It seems indeed to be possible to reach 
adiabatically any non-extremal state from any other one by 
passing through extremal states (which are non-isoareal). 
This would imply a failure of the second law of thermodynamics. 
In fact, it would be possible to find a Carnot cycle having 
thermal efficiency equal to one, against Ostwald's statement of the 
second law. In other terms, it would be possible to transform heat 
entirely into work, against the second law. 
On this topic, see a discussion in sect. \ref{cn}. Moreover, 
because of the intersection of integral manifolds, even if 
only at $T=0$, one cannot conclude that there is a foliation 
of the whole thermodynamic manifold [extremal manifold included] 
but what one could define an almost-foliation, i.e. a foliation 
except for a zero measure set (the integral manifold $T=0$).\\
From 
a physical point of view, in order to avoid the above 
singular behavior of the thermodynamic foliation, 
one could {\sl decide} that  
the surface $T=0$ should be a leaf itself, that is, to {\sl exclude}  
the set of solutions (\ref{nonextone}) for the above Cauchy problem.  
Notice that the set of solutions (\ref{nonextone}) corresponds 
to the isoareal solutions $dA=0$. Obviously, the existence of extremal 
black holes which have the same area as non-extremal states is not 
questioned; what is questioned by refusing the set of solutions 
(\ref{nonextone}) is the validity of $S\propto A$ also 
in the case of extremal black holes. 
The geometric foliation of the whole black hole manifold whose leaves 
are given by $A=$ const. should correspond 
to the thermodynamic foliation only in the case of non-extremal states.  

In order to avoid problems with thermodynamics, 
one could construct a foliation of the thermodynamic manifold 
whose leaves are 
\beqnl
&&\hbox{the\ surfaces}\ S=A/4=\hbox{const.\ for non-extremal states}\\ 
&&\hbox{the\ surface of extremal states}.
\label{folia}
\eeqnl
The leaves for the non-extremal manifold are the usual ones, which 
can be generated by means of the Pfaffian form $\delta Q_{rev}$. Instead, 
the exceptional integral manifold $T=0$ is {\sl assumed} to be isentropic 
and such that any extremal state has an entropy which is different 
from the one of any near extremal state (or, by continuity, one could 
pass adiabatically from non-extremal states to extremal ones).  
The Bekenstein-Hawking law ensures that $S=A/4$ can assume 
arbitrarily large values and, a priori, also very small values, 
the only lower bound could be given by the onset of a 
quantum gravity regime (which occurs in the limit $M\to 0$). 
Without considering such small values of $S$ implying a quantum 
gravity regime, 
it is reasonable, on a purely thermodynamic footing, to 
assume that $S_E=0$ for any extremal state. A further discussion 
is found in the following.\\
As a consequence of the latter assumption, the thermodynamic foliation 
one obtains is given by the following discontinuous entropy $S(M,Q,J)$:
\beqnl
S(M,Q,J) &=&
\frac{A}{4}\quad \hbox{for non-extremal states}\\ 
&=& 0\quad \hbox{for extremal states}.
\label{folias}
\eeqnl
About the thermodynamic consistency of this assumption, and 
its relation with (U), see also the discussion in sect. \ref{second-u} herein.

\subsection{The Bekenstein-Hawking formula 
for extremal states and the Carnot-Nernst cycle}
\label{cn}
 
\begin{figure}[h]
\setlength{\unitlength}{1.0mm}
\centerline{\psfig{figure=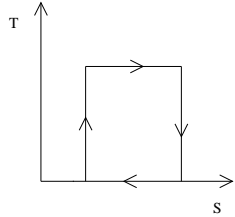,height=3cm,angle=-90}}
\vspace{0.2cm}
\caption{Carnot--Nernst cycle involving the isotherm at $T=0$. 
Its efficiency is one, against the second law.}
\label{fig3}
\end{figure}

We discuss in detail the problems that can arise if the above failure 
of the adiabatic inaccessibility property is verified. 
Let us consider a thermal 
Carnot cycle in the plane $T-S$ (see. fig. \ref{fig3}; the cycle is    
clockwise), 
having the lower isotherm exactly at $T=0$. 
We define it as Carnot--Nernst cycle.
 
If it is possible to perform a Carnot-Nernst cycle, 
then one can be able to construct a thermal machine with 
efficiency exactly equal to one, in violation of the 
second law of thermodynamics. This argument is substantially due to 
Nernst, who introduced it for supporting the third law of thermodynamics 
(see e.g.   \cite{landsberg,landrmp,landstat}). 

Of course, one has to check the actual possibility to perform 
such a cycle. 
The following 
hypotheses are taken into consideration:\\
\\
$\alpha$) extremal states can be reached by means of 
reversible adiabatic paths by starting from non-extremal states;\\ 
$\beta$) reversible transformations along extremal states 
discussed in sect. \ref{ebh} are allowed;\\
$\gamma$) non--isoareal transformations along extremal states 
exist;\\  
$\delta$) the Bekenstein--Hawking law holds for extremal states.\\
\\
If the hypotheses $\alpha),\beta),\gamma)$ and $\delta$)  
are all verified, then the Carnot-Nernst cycle can be performed and 
a violation of the second law occurs. 
We discuss now the above hypotheses.

Hypothesis $\alpha$) is verified, in the light of the 
existence of integral manifolds of $\delta Q_{rev}$  
which are allowed to reach $T=0$, unless some discontinuity 
occurs.

Hypothesis $\beta$) is more critical. In the case of standard 
thermodynamics objections against the possibility to 
perform a reversible transformation at $T=0$ have been raised \cite{pippard}, 
because of the impossibility to improve a change between the adiabatic 
constraint used in approaching $T=0$ and the adiabatic constraint 
in performing the adiabatic isotherm $T=0$. 
The possibility to perform of transformations along $T=0$ states 
in standard thermodynamics 
has been criticized by Einstein \cite{einstein}  
(cf. also \cite{simon,landrmp}), 
both from the point of view of the unavoidable presence of 
non-negligible irreversibility occurring near the absolute zero, 
and from the point of view of the actual possibility to perform 
an ideal transformation along $T=0$. 
Actually, e.g. in the case of a Reissner-Nordstr\"{o}m black hole, 
one could effectively distinguish between the adiabat approaching 
the extremal states (which is characterized by the equation $Q^2=2 M r_+ 
-(r_+)^2$, where $r_+$ is the radius of the initial black hole state), 
and the adiabat along extremal states (whose equation is $Q^2=M^2$). 
Rejecting a priori $\beta$) would mean implicitly to introduce 
a  ``discontinuity'' for thermodynamics in the 
behavior of extremal states with respect to non-extremal ones. 
The impossibility to perform  any transformation along the $T=0$ 
isothermal surface would be surely peculiar, in the sense that 
it is a property which distinguishes the $T=0$ submanifold 
with respect to the thermodynamic space at $T>0$.   
\\
Moreover, one has to ensure that transformations such that $dA<0$ 
along the extremal submanifold can occur in order to 
implement the Carnot-Nernst cycle in black hole thermodynamics. 
We have found no definitive counter-example against $dA<0$ 
along extremal states. 

Hypothesis $\gamma$) is introduced because the existence of 
non-isoareal transformations along extremal states is required 
if one has to perform the Carnot-Nernst cycle. $\gamma$) 
is verified, because  
$dA\not =0$ is allowed along extremal states (see sect. \ref{ebh}). 

If we assume that also hypothesis $\delta$) holds, then 
then a violation of the second law in a thermostatic 
framework occurs. As a consequence, 
we look for solutions. Moreover, whichever doubt one could raise 
against the possibility to perform the Carnot-Nernst cycle, 
one can agree with Einstein's statement that 
the existence of adiabatic paths allowing a Carnot-Nernst cycle 
is ``very hurtful to one's physical sensibilities'' \cite{einstein}.

A possibility to avoid the failure of the second law implied by the 
Carnot-Nernst cycle consists in rejecting $\alpha$) in the frame 
of black hole thermodynamics, by requiring that a 
discontinuity does not allow to perform the adiabats and reach 
the extremal states (even if non-extremal states are 
dense  near the extremal ones along adiabats).
\footnote{In the frame of standard thermodynamics, Nernst 
invokes the failure of hypothesis $\alpha$) in the sense of the 
absence of adiabats reaching $T=0$, whereas 
in black hole thermodynamics one can invoke (U) only in order 
to protect the second law (obviously 
the equivalence between (U) and (N) is violated). 
In   \cite{landsberg,landstat}, the Carnot--Nernst cycle is utilized 
with the aim to justify the following principle: 
The zero temperature states are so poorly populated that it is impossible 
to draw a continuous line between them. This point of view 
introduces actually an element of topological difference between 
states at $T>0$ and states at $T=0$, a ``discontinuity'' in thermodynamics, 
unless one does not postulates that the density of states starts 
decreasing at a positive $T$ near enough to the absolute zero 
(this requirement could match Munster's statements about the 
failure of thermodynamics for finite-size systems sufficiently 
near $T=0$ \cite{munster2}). 
This poor population density does not seem to affect 
extremal black holes.} 
Notice that the failure of the Bekenstein-Hawking law for extremal states 
implements such an impossibility to attain along isentropes the 
extremal states. Then a failure of hypothesis $\delta$) implies a 
failure of $\alpha$), and  
the failure of hypothesis $\delta$) is a natural solution 
to the above problem. We show in the following subsection that the 
failure of $\delta$) is less unnatural than a first look could 
suggest, because a discontinuous behavior of thermodynamics is 
in any case verified.

\subsection{Discontinuity between extremal states and non extremal ones.  
a thermostatic frame analysis and the failure of $S=A/4$}
\label{extrdisc}

We show that, both if the Bekenstein-Hawking 
law is verified and if it is not verified along extremal states, 
one is forced to admit that thermodynamics does not behave 
continuously in passing from non-extremal states to extremal ones. 
If the  Bekenstein--Hawking law is violated, the discontinuity 
is evident. If, instead, the 
Bekenstein--Hawking law is maintained also in the extremal case, 
it is anyway true that a discontinuity is verified, because 
reversible transformations along extremal states are always adiabatic, 
as seen in sect. \ref{ebh},  
and in general non-isoareal (cf. also Appendix A), 
contrarily to what happens for non-extremal states, where 
adiabatic reversible transformations are necessarily isoareal. 

\begin{figure}[h]
\setlength{\unitlength}{1.0mm}
\centerline{\psfig{figure=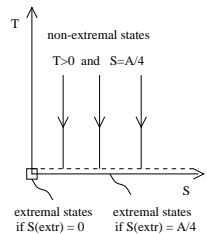,height=3cm,angle=0}}
\vspace{0.2cm}
\caption{$T-S$ plane in black hole thermodynamics. 
Three isentropes approaching extremal states are shown. 
The dashed line 
near $T=0$ indicates that a discontinuity in thermodynamics  
appears both in the case that $S=A/4$ holds for 
extremal states and in the case a different law is implemented 
(the case $S=0$ is displayed by means of a small box). 
}
\label{fig4}
\end{figure}

In the light of the hypothesis of failure for the Bekenstein-Hawking 
law along extremal states, we discuss shortly what happens from a 
statistical mechanical point of view. 
Extremal black hole entropy can be, in line of 
principle, valued by means of quantum mechanics: The Von Neumann 
entropy of extremal black holes is a priori calculable and it is 
the only meaningful entropy that can be associated with a state by  
studying it at exactly $T=0$, without considering a 
limiting process as $T\to 0^+$. 
A dichotomy between the limit as $T\to 0^+$ of the thermodynamic entropy 
and the $T=0$  
entropy for extremal states appears in   \cite{mitra} and 
in a paper concerning 
the quantization of extremal Reissner--Nordstr\"{o}m black holes 
\cite{louko}. Zero 
entropy is found by working separately on extremal states, a 
non-vanishing entropy is allowed if a limiting process starting 
from a quantization of non--extremal states is given rise\cite{louko}. 
The same dichotomy is implicit in   \cite{hawross,vanzo}, where $S=0$ for 
extremal states and $S=A/4$ for any non--extremal black hole
\footnote{In statistical mechanics, there are subtleties related 
with the order in which the limit as $T\to 0^+$ and the thermodynamic 
limit are taken in the calculation; the correct procedure consists 
in taking first the thermodynamic limit and then the limit as $T\to 0^+$.  
This further 
problem does not seem to be relevant for black holes.}. 
Doubts against limiting processes for calculating the 
entropy of extremal black holes 
are raised also in   \cite{louko} (therein and in   \cite{mitra} 
interesting comments about 
the results obtained for BPS states in string theory approach 
are found too). 
In black hole thermodynamics, one is suggested to introduce 
a discontinuity of thermostatics 
between $T>0$ and the absolute zero. 
A discontinuity is appealing in black hole thermodynamics 
because a very different physics is involved in the case of extremal black 
holes with respect to non-extremal ones (see e.g. 
  \cite{hawross,vanzo,bekenstein,dadh,racz}). 
From this point of view, one has also to take into account 
that, although a continuous behavior of some geometrical 
properties is verified, there are important differences in properties like 
the topology of the manifold. The Euler characteristic changes 
and this topological difference has been related with the 
thermodynamic differences between extremal and non--extremal 
black holes, being the global thermodynamic functionals
linked with the global properties of the manifold 
\cite{hawross,hawsurf,gibkal,libpol,wang}. 

In the following section, we relate in a thermostatic framework 
the unattainability property and the failure of the Bekenstein-Hawking law.

\section{Thermostatics, (U) and $S_E=0$} 
\label{thermouna}

The violation of (N) near the extremal states $M<+\infty$ from a thermodynamic 
point of view does not forbid the attainment of the zero temperature state. 
In order to conciliate the validity of (U) and the violation of 
(N) a reasonable hypothesis is  
that near $T=0$ in black hole thermodynamics a (possibly abrupt) change in 
thermodynamic properties of the system occurs. We are inspired by Landsberg's 
hypothesis c), relative to a possible discontinuity ensuring (U) against the 
failure of (N). It is very interesting, because in studying the 
implication (U)$\Rightarrow$(N) Landsberg not only 
postulates the validity of (U) but also he tries to 
retrace the possibility to get (U) and not (N) in a peculiar behavior of 
some thermodynamic functions. The infinite mass case shows that 
it is not strictly necessary such a behavior, because the attainment of 
$T=0$ is forbidden in that case simply by the first law (conservation 
of energy). In standard thermodynamics, when (N) holds, the attainment 
of the absolute zero is generally thought to be forbidden by the 
second law (impossibility of $\Delta S<0$ for an adiabatic process of a 
closed system). 
For the finite mass extremal case we have shown that there is 
the possibility  to de--link (U) from (N) but we think it is 
interesting to investigate also if there are thermodynamic arguments 
suggesting (U) beyond the dynamic theorem of Israel \cite{israel}, whose 
thermodynamic implications are discussed in sect. \ref{irre}. 
It is also useful to recall the potential relevance of (U) in relation with 
the Cosmic Censorship Conjecture (CCC), the third law in the (U) 
form appears as a sort of ``thermodynamic side" of the CCC   
\cite{davies,israel,lousto,carter,sciama,roman,farrugia,proszy,sullivan}. 
See also \cite{parikh}.

In order to corroborate the hypothesis of a discontinuity in the 
sense of Landsberg, 
the most straightforward study involves the analysis of 
the behavior of ``standard'' adiabatic quantities near the extremal 
limit. If e.g. some adiabatic compressibility should vanish then 
the hypothesis would be verified, indeed it would be impossible to 
carry out the adiabat connecting a non--extremal state to an extremal one. 
Landsberg makes 
the example of an abrupt divergence in the elastic constants of a 
solid as a conceivable ideal process preventing a solid to reach 
a zero temperature state by means of quasi--static 
adiabatic volume variations (the hypothesis 
of   \cite{landsberg} is compatible with the vanishing near 
$T=0$ of the (adiabatic) compressibilities that are related 
with elastic constants 
in ordinary thermodynamics; particularly, for standard systems one can 
define the compressibility modulus as the inverse of the compressibility; 
it is proportional to the Young modulus in the case of a solid). 
But our analysis does not show a peculiar behavior of adiabatic 
derivatives and does not suggest the kind of discontinuity  characterizing 
Landsberg's hypothesis. Cf. also the appendix of   \cite{tranah}, where 
some adiabatic derivatives are calculated.\footnote{Nevertheless, 
in a thermostatic frame, 
the analysis of stability properties for Kerr--Newman black holes 
can suggest the (U) property of 
extremal states and are carried out in \ref{unsta}.}

We now give a 
thermostatic argument in favor of the unattainability 
of extremal states. It is based on 
the assumption that extremal states have zero entropy. 

\subsection{(U) and extremal states with $S_{E}=0$}
\label{second-u}
 
It is remarkable that, if extremal states have $S_{E}=0$, then 
one has a natural naive argument for implementing (U). 
Let us consider the following hypotheses:\\ 
\\
s1) the system is composed by a non--extremal black hole 
and matter; it is insulated, i.e., no exchange of energy, both in 
form of heat and in form of work, is allowed.\footnote{For the principle of 
increase of entropy, it would be sufficient that the system is 
closed and undergoes an adiabatic transformation.}\\ 
s2) Both the initial and the final state are 
equilibrium states (which can be a non-trivial postulate 
for states at $T=0$).\\ 
s3) The final state is an equilibrium state between the 
black hole and matter.\\
s4) $S_{E}=0$.\\
\\
Hypothesis s1) means that the principle of increase of entropy holds 
for the system. It implies that energy  
exchanges with the rest of the universe are forbidden. 
In a collapse situation, 
there can be surroundings of the system matter$+$black hole 
whose entropy variation could play a role in the application 
of the second law of thermodynamics. 
Hypothesis s2) is natural, in the sense 
that the thermodynamic entropy is unambiguously defined only 
for equilibrium states. 
s3) and s4) are discussed below. 
We can consider at first the case in which 
all the matter can be used for making extreme the 
black hole. For the initial 
state one has contributions to the entropy from the matter and 
a non-extremal black hole (NE): 
$S^{in}_{tot}=S^{in}_{NE}+S^{in}_{mat}$; in the final state 
one has only an extremal black hole, so that, 
according to s4),  $S^{fin}_{tot}=S_E=0$. 
Clearly the second law requires for the adiabatic process 
$S^{fin}_{tot}\geq S^{in}_{tot}$, which is impossible in our case.\\ 
One can also relax the hypothesis that all the matter is 
used for making extreme the black hole; if the matter at the end is 
in equilibrium with the black hole according to 
hypothesis s3), then the second law would 
require  $S^{fin}_{tot}=S^{fin}_{mat}\geq S^{in}_{NE}+S^{in}_{mat}$, 
which for ordinary matter is again impossible ($S^{fin}_{mat}\sim 0$ 
because of (N)). One could then arbitrarily approach an extremal state 
but the ``jump'' {\sl onto} extremality would be forbidden by the 
second law.\\ 
The final state could also be an equilibrium state 
if at least a portion of the residual matter is kept 
thermally insulated with respect to the black hole (violation 
of condition s3)) by some external mean,  
in which case the final state should have a 
contribution $S^{fin}_{mat}>0$. But it is hard to see how this 
framework could allow the attainability of extremal states without a 
violation of the second law. One should allow the 
formation of extremal states from non-extremal ones in 
such a way to preserve the principle of increase of the entropy 
for the thermodynamic universe under consideration\footnote{For 
``thermodynamic universe'' we mean the smallest closed and 
thermally isolated system of interest (e.g., in the case of a 
black hole and matter falling into it, if thermal and matter exchanges 
with the surroundings are impossible, the thermodynamic universe 
is the system black hole $+$ matter).} 
(an highly non-trivial task in light of the 
fact that $S^{in}_{NE}$ is a huge number in 
general).

This hypothesis relating (U) to the second law is to be compared with 
processes that allow to get extremal states. In general, they 
don't correspond to quasi-static processes. 
Extremal black hole formation from extremal collapsing thin shells 
\cite{vanzo} and from charged thin shell collapsing on 
a non--extremal black hole (see   \cite{farrugia}) are 
examples of these processes. Could one   
define their initial state as an equilibrium state? In the first 
example, the shell is pushed from infinity.   
In the second case, the shell has to be fired onto the non-extremal 
black hole. 
If the quantum gravity result $S_{E}=0$ is true, then 
a careful analysis of the second law is required in order 
to ensure that, at a deeper level with respect to the naive 
analysis for the aforementioned processes, the second law is 
actually preserved. Even if this 
analysis were essentially unmodified,  
a detailed 
analysis of stability could reveal that the probability of 
these processes is very low; in   \cite{proszy} is indeed 
underlined that the extremalization process by means of 
thin shells is highly unstable under perturbations (see the 
conclusions therein). Quantum effect could also play 
a relevant role in this case, as follows from 
  \cite{anderson,sonego}. In   \cite{hod},
it is conjectured $S_E=0$ by starting from quantum 
gravity considerations. It is interesting to note that 
there is consistency with our conjecture for implementing (U) in 
a thermostatic context.    

\subsection{$S_E=0$ and the merging of two extremal black holes}

We discuss a consistency check for the hypothesis $S_E=0$. Let us consider 
two black extremal black holes $(M_1,Q_1,J_1)$ and 
$(M_2,Q_2,J_2)$ (one variable, e.g. $M$, of course depends on the 
other two because of the extremal constraint). Let us allow the 
two extremal black holes to merge and that no energy is exchanged with 
the rest of the universe during the process [one could as well consider 
a more general situation in which the couple is thermally insulated, i.e., 
it does not exchange heat with its surroundings]. 
We suppose that the final state consists of a single black hole resulting 
from the merging of the two initial extremal states. 
We wonder if the final state could be extremal. The point is 
that, if the final state could be extremal, then a violation of the 
second law could still occur, in fact the process is irreversible and 
the final entropy should be greater than the initial one but, 
if the final state is extremal and no energy 
is exchanged with the rest of the universe,  
one would find $S_{E1}+S_{E2}=S_{in}=0=S_{E12}=S_{fin}$. 
Let us define, as in   \cite{tranah}, 
\beq
a^2(M,Q,J)\equiv M^4-M^2 Q^2-J^2;
\eeq
from   \cite{tranah} we know that, by defining
$a_{12}^2\equiv a^2(M_1+M_2,Q_1+Q_2,J_1+J_2)$,\\  
$a_1^2\equiv a^2(M_1,Q_1,J_1)$,  and $a_2^2\equiv a^2(M_2,Q_2,J_2)$ 
one has 
\beqnl
a_{12}^2-(a_{1}+a_{2})^2 &=& 
\left( \frac{M_2}{M_1} a_1-\frac{M_1}{M_2} a_2 \right)^2+
\left( \frac{M_2}{M_1} J_1-\frac{M_1}{M_2} J_2 \right)^2\cr
&+&2 M_1 M_2 \left[\left(M_1+M_2\right)^2-\left(Q_1+Q_2\right)^2\right]\cr 
&+&
2 \left(M_1 M_2-Q_1 Q_2\right) \left(M^2_1+M^2_2\right).
\eeqnl
We wish to see if it is possible that $a_{12}=0$, which would imply 
that the final state is extremal. In our case, 
the above formula simplifies because $a_1=0=a_2$ for the initial 
extremal states. Moreover, one has $M_1=M_{1E}$ and $M_2=M_{2E}$. 
Then one finds
\beqnl
a_{12}^2 &=& 
\left( \frac{M_{2E}}{M_{1E}} J_1-\frac{M_{1E}}{M_{2E}} J_2 \right)^2\cr
&+&2 M_{1E} M_{2E} \left[\left(M_{1E}+M_{2E}\right)^2-
\left(Q_1+Q_2\right)^2\right]\cr 
&+&
2 \left(M_{1E} M_{2E}-Q_1 Q_2\right) \left(M^2_{1E}+M^2_{2E}\right).
\eeqnl
If $J_1\not =0$ and/or $J_2\not =0$, no matter which values one 
considers for $Q_1,Q_2$,  
then $a_{12}>0$ and the 
final state is a non-extremal state. The final entropy is surely 
much greater than the initial one. If $J_1=0=J_2$, then it is possible to 
find a final state which is still extremal if one merges two 
extremal Reissner-Nordsstr\"{o}m black holes having charges with 
the same sign, as it is evident from 
\beqnl
a_{12}^2 &=&
2 M_{1E} M_{2E} [(M_{1E}+M_{2E})^2-(Q_1+Q_2)^2]\cr 
&+&
2 (M_{1E} M_{2E}-Q_1 Q_2) (M^2_{1E}+M^2_{2E}).
\eeqnl
One should question 
if it is possible to allow $S_E=0$ and to preserve 
the second law in a real process of merging. Notice that even 
a very small angular momentum would protect the second law.\\ 
\\
We don't study here this problem, we limit ourselves 
to the above considerations. Of course, in light of the 
risk for violations of the second law, also the 
hypothesis s4), $S_E=0$ should be questioned.

\section{Irreversibility Frame}
\label{irre}

We now discuss the meaning of (U) as it is rigorously 
proved in   \cite{israel}.  
It is important to stress that in Israel's proof  
(U) holds from a dynamic point of view. 
In particular, it is shown that a non-extremal black hole cannot 
become extremal in a finite advanced time if the accreted matter 
stress--energy tensor satisfies the weak energy condition in 
a neighborhood of the 
outer apparent horizon and remains bounded and continuous \cite{israel}. 
Israel's result shows that, 
along a continuous process, in a finite advanced time 
it is impossible to destroy the trapped surfaces, which are present 
in the non--extremal states and instead are missing in the extremal one. 
This implies that the 
process (NE)$\to$(E) from a non--extremal black hole $r^{NE}_{+}-r^{NE}_{-}>0 
\Leftrightarrow k^{NE}>0$ to an extremal black hole 
$r^{E}_{+}-r^{E}_{-}=0 \Leftrightarrow k^{E}=0$ as a final state 
requires an infinite time\footnote{The 
process (NE)$_1 \to$(NE)$_2$ between non extremal states (NE)$_1$ 
and (NE)$_2$ can occur in a finite time.}. 
In the following we stress that 
also the subclass of thermodynamic processes 
is constrained by Israel's result, at least as far as 
{\sl approximations of quasi--static processes are implemented 
by means of accretion of matter whose stress--energy tensor is 
bounded and continuous and satisfies the hypotheses of   \cite{israel}.} 
Approximations of quasi-static processes e.g. by means of 
point-like particles satisfy the aforementioned requirements if 
suitably corrected (they would imply a distributional stress--energy 
tensor that should be corrected by taking into account the finiteness of 
the Compton wavelength of the particles).  
Some dynamic restrictions in Israel's proof  
don't allow a full identification of such a result with 
the unattainability (U4) of the extremal states.\\

We show that Israel's result implies that extremal black holes 
cannot be considered as equilibrium states contiguous to non--extremal 
black hole equilibrium states. The thermodynamic manifold of equilibrium 
states has to present a discontinuity.\\ 
The framework of irreversible 
thermodynamics is the most appropriate in order to include in thermodynamics  
Israel's dynamic information. In fact, irreversible thermodynamics 
allows the introduction of the notion of 
relaxation time to an equilibrium state, 
and gives the effective physical time-scale with respect to which a 
process can be considered properly as a good approximation of a 
quasi--static process.  
The inaccessibility in a 
finite time of the extremal states from non--extremal ones 
by means of a continuous process can be rephrased as the impossibility 
to carry on an approximate quasi--static process joining 
non--extremal states to extremal 
ones, due to the divergence, which is met in approaching extremal states, 
of the ``relaxation'' time to the equilibrium state (that corresponds 
to the formation time of the stationary black hole state). 
We corroborate this point of view as follows. In   \cite{israel} 
is found that, in order to squeeze out trapped surfaces, it is necessary 
an infinite time $\tau_{no\ trapped}=\infty$. Since an extremal 
black hole has no trapped surfaces, it follows the third principle 
$\tau_{(NE)\to (E)}=\infty$. In the framework of irreversible 
black hole thermodynamics one can conclude that 
$\tau_{relaxation}$ to an extremal equilibrium state coincides 
with $\tau_{(NE)\to (E)}=\infty$ if the latter is relative to 
a generic dynamical bounded and continuous process as in   \cite{israel}, 
even if a correspondence 
between relaxation phenomena and irreversible 
black hole thermodynamics still is missing. 
See however   \cite{cartergen}, 
in particular section 6.3.3 therein, where a 
time-scale $\tau \sim 1/T_{\mathrm bh}$ 
is proposed for the decay of a perturbed black hole to a stationary state  
\cite{cartergen}\footnote{Note that 
also the ``mining process'' time-scale of   \cite{roman} is of the same 
order.}. Aspects of irreversible black hole thermodynamics are also 
explored in   \cite{damogros} (where a formation time-scale again order of 
$1/T_{\mathrm bh}$ appears) 
and in particular in   \cite{sciamait,sciacand}. We don't develop 
herein an irreversible thermodynamics formalism for black holes, which  
should be the subject of further investigations.

Israel's result implies that thermodynamic formalism involving 
quasi--static 
processes cannot be extended to the extremal states because, from a 
physical point of view, approximating ideal quasi--static processes 
by means of (roughly) very slow processes 
looses sense in processes involving a transition from near extremal states 
to extremal ones because of the infinity in the 
relaxation time. Equilibrium thermodynamic  
formalism gives rise to a consistent description of 
very slow processes only   
if the relaxation times 
of various parameters defining the equilibrium state are much bigger 
or much smaller than the measurement time. In the former case the 
the parameters get a constant value, in the latter they get their 
equilibrium value \cite{munster}. The case of a measurement time 
of the same order as the 
relaxation time is critical \cite{munster}. In the case of 
a black hole, a measurement time suitable in order to measure the 
reaching of a black hole equilibrium state should be much longer than 
the formation time. For the case of an extremal black hole state 
attained by means of a continuous process, 
there is no satisfactory measurement time because at best an infinite 
measurement time should be compared with an infinite relaxation (formation) 
time. This means also that there is an intrinsic inaccessibility of extremal 
states if they have to be reached by means of a 
continuous quasi--static process. 
Other considerations about the failure 
of thermodynamics near the extremal limit can be found 
in   \cite{preskill}.

We can now implement a better comparison of Landsberg's 
hypothesis c) with the actual behavior of black holes near the extremal 
limit as dictated by   \cite{israel}. 
The infinity of the time required in order 
to get an extremal state suggests that the dynamic process meets some 
hindrance near the extremal state to be carried further on in a finite time, 
if the process 
is ``continuous''. Qualitatively, this is e.g. suggested by the fact that  
adding charge to a non--extremal Reissner--Nordstr\"{o}m black hole 
becomes more and more difficult 
due to the increase in the electrostatic repulsion. 
The impossibility to increase the black hole charge/angular momentum 
in a finite time till the extremality condition is implemented 
just resembles Landsberg's suggestion \cite{landsberg} of a discontinuity 
near the absolute zero (violation also of the hypothesis c)), but
it involves an intrinsically dynamical information (as the infinite 
time required 
in order to implement the process). 
The peculiar characteristic of black hole thermodynamics, 
from the point of view of the third principle, is that 
``infinity of processes'' 
of the standard formulation for (U) is substituted with ``infinity of time''. 
Actually the latter formulation seems more general than the former, 
in the sense that ``infinity of processes'' can easily imply   
``infinity of time'' (but obviously not vice versa). 

A final comment can be made concerning (U) according to 
Israel and (U) as required by the second law in \ref{second-u}. 
Let us one 
assumes the following conjecture: {\sl The gravitational entropy of black 
holes is non--zero and it is given by Bekenstein--Hawking law only in 
presence of trapped surfaces}. Then Israel's proof can be interpreted 
as the dynamics-based side of the thermodynamics-based  
hindrance to reach extremal states shown in \ref{second-u}.  
It is also tempting to remark that the aforementioned conjecture, 
together with the above thermodynamic analysis, can  
enforce the conjecture  interpreting black hole 
entropy as ``entanglement entropy'' \cite{bkls}. In fact, 
the ``entanglement entropy'', which is obtained by 
tracing the von Neumann entropy over unavailable degrees of freedom, 
can be associated with a trapped surface (the trace should be 
taken over the quantum field modes contained in the trapped surface).  
Moreover, for an ``entanglement entropy'' satisfying the area law 
there is no reason why 
it should approach a constant (zero) value as the extremal states are 
approached\footnote{For the value of the 
constant see also   \cite{qotd}.}, 
because the area of the trapping surface depends on geometric 
parameters and it does not vanish near the extremal states, no matter how 
near the extremal boundary the black hole could be. 
Even the 
discontinuity of $S$ at the extremal boundary could be justified 
(no trapped surface would mean zero entanglement entropy).\\
We don't want to claim that ``entanglement entropy'' is 
mandatory, we simply limit ourselves to note that it seems to have 
chances to match both the failure of (N) and the validity of (U) in 
black hole thermodynamics. The alternative view to consider 
the failure of (N) as due to ``frozen non-equilibrium states'' in 
black hole thermodynamics does not seem to be plausible. 
See a discussion in Appendix B.

\section{Conclusions}

We have analyzed the third law of thermodynamics from the 
point of view of a purely thermodynamic framework. A particular 
reference to P.T.Landsberg ideas has been made. 
We have discussed both the branches as $T\to 0^+$ one can find 
in thermodynamics. 
An analysis of the failure of the implication (U)$\Rightarrow$(N) 
in black hole thermodynamics has revealed  a special 
status to (U) in black hole thermodynamics. 
The analysis of the limit $T\to 0^+$ for large black hole 
masses, where (U) is necessarily implemented,  has shown that (N) is 
satisfied if the black hole is uncharged. 
In the branch $M\to \infty$ the failure 
of (N) vs. the validity of (U4) in the charged case has to be 
considered as a particular property of black hole thermodynamics.\\
In the framework of 
Carath\'eodory's approach to thermodynamics, we have point out 
which kind of problems can arise if the Bekenstein-Hawking 
law holds also for extremal states. In particular, the possibility to 
get a Carnot-Nernst cycle has been pointed out. Being such a cycle 
involved in a violation of the second law of thermodynamics, 
we have discussed its performability in the framework of 
black hole thermodynamics. The violation of the Bekenstein-Hawking 
law has been suggested as a viable  
solution for this problem. Moreover, 
we have shown that, by requiring that the entropy of extremal black 
holes is zero, one can support (U) from a thermostatic point of 
view. We remark that  
finding e.g. $S=0$ for extremal states does not mean that (N) is 
valid, unless such a result is corroborated by a limit approach 
as $T\to 0^{+}$. In fact (N) gets, as a matter of facts, its real 
meaning and its real experimental verifications in standard 
thermodynamics if it is intended 
not as the behavior of the entropy at exactly 
$T=0$, but as the limit of entropy differences as $T\to 0^{+}$.\\
The result of 
  \cite{israel} (the appropriate frame 
is a dynamic one and an irreversible thermodynamic one) 
has been interpreted as a strong 
corroboration for the picture in which  
extremal states are separated by a discontinuity 
with respect to non-extremal ones. Qualitative arguments allowing 
to match Israel's theorem both with the hypothesis 
$S_E=0$ and with the interpretation 
of black hole entropy as entanglement entropy have also been introduced.

\section*{Acknowledgments}

FB wishes to thank C.Destri and A.Parola for interesting 
discussions at the preliminary stages of this work. 
Thanks are also addressed by FB to S.Liberati for his 
interesting comments on a old version of this 
paper.

\appendix

\section{Adiabatic transformations in black hole thermodynamics}

We determine the 
equations for reversible adiabatic transformations in black hole 
thermodynamics. 
In the non--extremal cases they correspond to 
isoareal transformations because of the Bekenstein-Hawking law. 

Adiabatic transformations for non--extremal 
black holes satisfy the equation 
$$
A=A_0=\hbox{const.}
$$
that in the general case of a Kerr-Newman black hole becomes
$$
r_{+}^2+\frac{J^2}{M^2}=C
$$
where $C=A_0/(4 \pi)$ is a positive constant. 
If one defines $x\equiv M^2; 
y\equiv Q^2; z\equiv J^2$ then the above equation is equivalent 
to the following one:
$$
2 x-y+2 \sqrt{x} \sqrt{x-\frac{z}{x}-y}=C.
$$
One can solve e.g. for $z$ and find
$$
z=C x-\frac{1}{4}(y+C)^2.
$$
It is easy to show that $x\in (x_L^2/C,x_L)$, where 
$x_L\equiv (y+C)/2$. The extremal sub-manifold is 
defined by
$$
z_{E}=x^2-x y
$$
and its intersection with the above isoareal surface takes place at 
\beq
x-\frac{1}{2}(y+C)=0\ \Leftrightarrow\ x = x_L.
\label{tang}
\eeq
The extremal state surface and the 
isoareal surface are tangent. In fact, their 
tangent planes coincide along (\ref{tang}).\\

We can conclude that there are two classes 
of adiabatic transformations: 1) standard adiabatic transformations 
for non--extremal black 
holes; they are isoareal and isentropic; 2) ``extremal" 
adiabatic transformations that in general are not isoareal. 
In the 
Kerr-Newman extremal case it is possible to allow also for isoareal 
extremal transformations that can be obtained by imposing 
$dA_{E}=0$. Their equation is 
$$
Q^4+4 J^2=(\frac{A_0}{4\pi})^2     
$$
They represent a nontrivial sub-manifold of the general extremal 
case. If $J=0$ or $Q=0$ one gets that this manifold becomes 
a single point.

\section{Glassy systems, frozen equilibrium and (N)}

It is well known that a wide discussion 
about the validity of (N) was given rise by some physical systems 
that seemed to violate Nernst's postulate of isentropy of the 
zero--temperature states. Actually, it has been shown that these 
peculiar systems (e.g. CO and glassy substances) don't satisfy the 
condition of internal equilibrium, 
that is, near $T=0$ some degrees of freedom remain frozen in a 
non--equilibrium meta--stable configuration\cite{simon,guggenheim,wilson}. 
Elements of configurational disorder can remain unchanged during 
the cooling down of the system towards a low temperature. 
The relaxation time 
to a condition of inner equilibrium is much bigger than the measurement 
time and it can be effectively 
infinite. A residual molar entropy is then allowed at $T=0$. 
Long-time measurements have shown the convergence to (N) of the 
calorimetric entropy for some substances violating (N). 
Thus, a further hypothesis has been added in order to 
ensure the validity of (N), which is the condition of internal equilibrium 
\cite{simon,wilson}. There is no definitive agreement 
about this hypothesis.\\  

In the black hole case, it is still difficult to find out 
a definitive notion of ``internal states'' and of micro--states. 
``Meta-stability of non--equilibrium states'' and ``frozen-in 
disorder'' can hardly justify  the 
violation of (N) in the case of black holes. 
The difference with respect to the case 
of the apparent violation of (N) in ``glassy systems'' having 
finite relaxation times is evident, 
indeed by increasing the measurement time no convergence to 
the implementation of (N) can be expected for black holes.

\section{Failure of concavity and (U)$\Leftrightarrow$(N)}

In the following, we analyze what happens if condition a) is relaxed. 
In fact, in black hole thermodynamic a) is not satisfied.
For simplicity of notation, we substitute $S_{T,x}$ for 
$S(T,x^1,\ldots,x^n)$ and $C_x$ for $C_{x^1,\ldots,x^n}$.  

\subsection{Relaxing condition a) against (U)$\Rightarrow$(N)}

\begin{figure}[h]
\setlength{\unitlength}{1.0mm}
\centerline{\psfig{figure=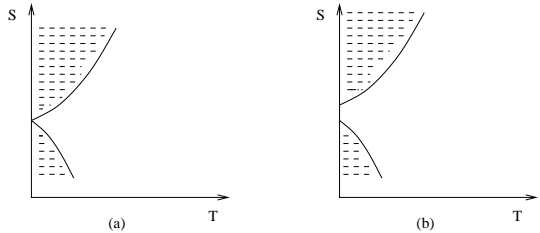,height=3cm,angle=0}}
\vspace{0.2cm}
\caption{ Isentropes approaching $T=0$ at finite parameters 
exist and (U) is ensured by a 
discontinuity in the sense of Landsberg. (N) is allowed only in case (a).}
\label{fig5}
\end{figure}

\begin{figure}[h]
\setlength{\unitlength}{1.0mm}
\centerline{\psfig{figure=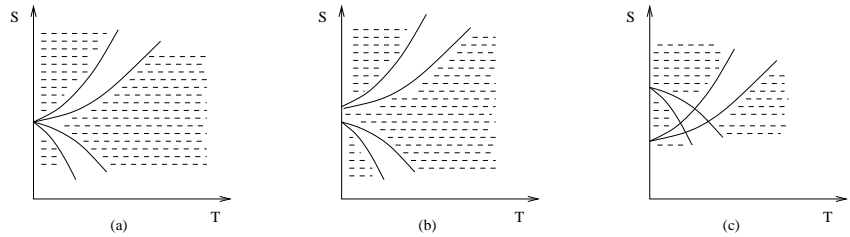,height=3cm,angle=0}}
\vspace{0.2cm}
\caption{A multi--branches structure is allowed. In all cases, 
(U) is ensured without invoking discontinuities near the absolute zero. 
In (a), in spite of the presence of $C_y<0$ connected to 
$T=0$, (N) holds. In (b) 
no isentrope is intersected both by curves reaching $T=0$ and 
having $C_x>0$ and by 
curves reaching $T=0$ with $C_y<0$. 
In (c), such an isentrope exists but no 
isentrope reaching the absolute zero is allowed. In (b) and in (c) the 
entropic version (N) is violated.}
\label{fig6}
\end{figure}

The presence of heat capacities with opposite sign can invalidate 
the proof of the double implication (U)$\Leftrightarrow$(N). 
Let us assume then that (U) is satisfied and that there exist 
isometric curves (i.e. isometric transformations) reaching 
$T=0$ such that some have $C_x>0$ and other 
$C_y<0$. Note that the presence of heat capacities $C_y<0$ at constant 
deformation parameters means the failure 
of the standard concavity properties of the entropy. 
Then, this non--uniformity of the sign of the heat capacities  
along transformation reaching the absolute zero allows to violate 
the implication (U)$\Rightarrow$(N), and, moreover, it 
seriously jeopardizes the 
identification of (U) with the absence of isentropic reaching $T=0$.  
See fig. \ref{fig5}. 
Let us first consider the case where along different isometric 
transformations with opposite signs of the corresponding heat 
capacities and starting from $T=0$ it is possible to reach 
the same isentropic surface. Let us define 
\beqna
S_{T_1,x}&=&S_{0,x}+\int_{0}^{T_1} \frac{C_x}{T} dT  > S_{0,x}\cr
S_{T_2,y}&=&S_{0,y}-\int_{0}^{T_2} \frac{|C_y|}{T} dT < S_{0,y}.
\eeqna
If $S_{T_1,x}=S_{T_2,y}$ and  
if $T_1,T_2>0$, the equality 
$$
S_{0}+\int_{0}^{T_1} \frac{C_x}{T} dT=
S_{0}-\int_{0}^{T_2} \frac{|C_y|}{T} dT
$$
is obviously impossible. The only possibility is that $T_1=0=T_2$, 
but, then, (U) cannot be implemented in general as ``absence of 
isentropic transformation reaching $T=0$'', except for a  
$S-T$ diagram of the type sketched in fig. \ref{fig6}. 

Note that this reasoning concerning the non-uniformity of signs 
of heat capacities for transformations connected to $T=0$ 
can be easily extended to the case where 
generic curves $\gamma^{0}$ which arrive at $T=0$ and having 
$C_{\gamma^{0}}<0$ are allowed. 

If, instead, no intersection 
to the same isentrope is possible, then, again a violation of (N) 
can occur.  
In the following figures, some possible diagrams 
are sketched. They imply a violation of Landsberg's hypothesis 
b) and/or of c).\\  
As we have shown, (N) can 
be allowed for only if the starting isentropic coincides with 
the zero--temperature one (cf. fig. below), but, if $S\geq 0$ is assumed, 
then at least Planck's restatement of (N) has to be violated. 

\subsection{Relaxing condition a) against (N)$\Rightarrow$(U)}

\begin{figure}[h]
\setlength{\unitlength}{1.0mm}
\centerline{\psfig{figure=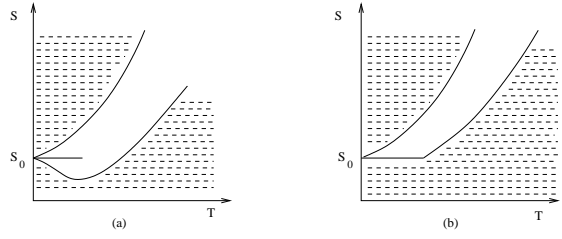,height=3cm,angle=0}}
\vspace{0.2cm}
\caption{Two examples of violation for the implication (N)$\Rightarrow$(U). 
(a) Paths having $C<0$ and reaching $T=0$ are allowed; (b) A behavior like 
the one of Wheeler's counter-example [58] is displayed (the original 
counter-example requires $S_0=0$).}
\label{fig7}
\end{figure}

It is also easy to deduce that the existence of paths with 
opposite sign near $T=0$ could  be reconciled with (N) without  
implying (U). In fact, if $C_x>0, C_y<0$ then 
\beqna
S_{0}&+&\int_{0}^{T_1} \frac{C_x}{T} dT > S_{0}\cr
S_{0}&-&\int_{0}^{T_2} \frac{|C_y|}{T} dT < S_{0}
\eeqna
and, in absence of suitable multi--branching (hypothesis b)),  
the isentropic $S=S_{0}$ allows to get $T=0$. In  
\cite{wheeler}
another counter-example to (N)$\Rightarrow$(U) is shown. See fig. \ref{fig7}. 
The system studied therein displays a 
particular behavior, in the sense that $S=0$ is attained at $T_1>0$ 
and $C_p=0$ for $0\leq T \leq T_1$ is allowed. Condition 
a) is then violated. 
According to standard proofs, $S=S_0$ cannot be attained at $T>0$. 

\section{Stability properties and (U)}
\label{unsta}

The entropy is 
a non--concave function even in the case in which the thermal stability 
is ensured (as near the extremal states). In fact, there are principal 
minors of the Hessian (see e.g.   \cite{tra2,callen}) 
that don't satisfy the concavity (stability) 
requirement. In   \cite{tranah} it is shown that in the black hole case 
the minor $\Delta_{3}$ does not implement the stability requirement 
for any value of the physical parameters. In fact \cite{tranah} 
$$
\Delta_{3}=\frac{\pi}{8} \frac{1+3 \Omega^2 Q^2-\Phi^2}{M T^5 S^3}
$$
and the stability requirement $\Delta_{3}<0$ is never satisfied. 
Moreover, the instability 
becomes maximal at the extremal limit (where all the minors diverge). 
We show explicitly this property in the Reissner--Nordstr\"{o}m and 
in the Kerr cases. Let us define 
\beq
a_{X_i X_j}\equiv \frac{\partial^2 S}{\partial X_i \partial X_j}
\eeq
where $X_i$ are the extensive variables appearing in the fundamental 
relation in the entropy representation. In the Reissner--Nordstr\"{o}m and 
in the Kerr cases we get respectively
\beqnl
a^{\mathrm RN}_{MM}&=&-(2 \pi) 
\frac{1}{(M^2-Q^2)^{3/2}} (M+\sqrt{M^2-Q^2})^2 (M-2\;\sqrt{M^2-Q^2})\cr 
a^{\mathrm RN}_{QQ}&=&-(2 \pi) 
\frac{1}{(M^2-Q^2)^{3/2}} (M^3+(M^2-Q^2)^{3/2})\cr                          
a^{\mathrm RN}_{MQ}&=&(2 \pi)^2 
\frac{Q^3}{(M^2-Q^2)^{3/2}}\\
&&\cr 
a^{\mathrm Kerr}_{MM}&=&(4 \pi) 
\frac{1}{(M^4-J^2)^{3/2}} (M^6-3 M^2 J^2+(M^4-J^2)^{3/2})\cr
a^{\mathrm Kerr}_{JJ}&=&-(2 \pi)  
\frac{M^4}{(M^4-J^2)^{3/2}}\cr 
a^{\mathrm Kerr}_{MJ}
&=&(4 \pi) \frac{J M^3}{(M^4-J^2)^{3/2}};
\eeqnl
and
\beqna
(a^{\mathrm RN}_{MQ})^2-a^{\mathrm RN}_{MM} a^{\mathrm RN}_{QQ}
&=& (2 \pi)^2 
\left(M^2-Q^2\right)^{-3/2}\cr 
&&\left[ M^3+(4 M^2-Q^2) \sqrt{M^2-Q^2}+
3 M^2 (M^2-Q^2) \right]\cr 
&&\cr
(a^{\mathrm Kerr}_{MJ})^2-a^{\mathrm Kerr}_{MM} a^{\mathrm Kerr}_{JJ}
&=& 8 \pi^2 M^4 
\left(M^4-J^2\right)^{-2} (M^2+\sqrt{M^4-J^2}).
\eeqna 
In the case of two independent thermodynamic variables $X_1,X_2$, stability 
requires that $a_{X_1 X_1}\leq 0, a_{X_2 X_2}\leq 0, (a_{X_1 X_2})^2-
a_{X_1 X_1} a_{X_2 X_2}\leq 0$ \cite{callen}. 
The third condition in both cases is always violated  
and moreover a divergence in the extremal limit appears. 
The condition that the entropy hypersurface lie everywhere below its family 
of tangent hyperplanes \cite{callen} is so violated maximally near the 
extremal states. Such an instability can give a thermodynamic reason for 
a thermodynamic ``runaway'' from extremal states and, as a consequence, 
also for (U).

\end{document}